\documentclass[trackchanges]{aastex701}

\usepackage{graphicx}
\usepackage{natbib}
\usepackage{multirow}
\usepackage{amsmath}
\usepackage{booktabs}
\begin{document}

\title{Rotation of a Solar Jet Driven by Plasma Flow along Helical Magnetic Fields in an Active Region}

\author[orcid=0009-0003-4609-3177,sname='Lei']{Lei Huang}
\affiliation{School of Astronomy and Space Science, Nanjing University, Nanjing 210023, People’s Republic of China}
\affiliation{Key Laboratory of Modern Astronomy and Astrophysics (Nanjing University), Ministry of Education, Nanjing 210023, People’s Republic of China}
\email{huanglei@smail.nju.edu.cn}  

\author[orcid=0000-0002-9293-8439,sname='Yang']{Yang Guo} 
\affiliation{School of Astronomy and Space Science, Nanjing University, Nanjing 210023, People’s Republic of China}
\affiliation{Key Laboratory of Modern Astronomy and Astrophysics (Nanjing University), Ministry of Education, Nanjing 210023, People’s Republic of China}
\email[show]{guoyang@nju.edu.cn}

\author{Zhen Li} 
\affiliation{School of Astronomy and Space Science, Nanjing University, Nanjing 210023, People’s Republic of China}
\affiliation{Key Laboratory of Modern Astronomy and Astrophysics (Nanjing University), Ministry of Education, Nanjing 210023, People’s Republic of China}
\email[show]{lizhen@nju.edu.cn}

\author[orcid=0000-0002-4205-5566,sname='Jinhan']{Jinhan Guo} 
\affiliation{School of Astronomy and Space Science, Nanjing University, Nanjing 210023, People’s Republic of China}
\affiliation{Key Laboratory of Modern Astronomy and Astrophysics (Nanjing University), Ministry of Education, Nanjing 210023, People’s Republic of China}
\email{jinhan.guo@nju.edu.cn}

\author[orcid=0000-0002-4978-4972,sname='Mingde']{Mingde Ding} 
\affiliation{School of Astronomy and Space Science, Nanjing University, Nanjing 210023, People’s Republic of China}
\affiliation{Key Laboratory of Modern Astronomy and Astrophysics (Nanjing University), Ministry of Education, Nanjing 210023, People’s Republic of China}
\email{dmd@nju.edu.cn}

\begin{abstract}
Solar jets, collimated plasma ejections driven by magnetic reconnection, play a vital role in energy transport and coronal heating. While rotational motions in jets are often attributed to magnetic field untwisting, alternative explanatory mechanisms remain possible. This study investigates a rotating jet in an active region observed on 2023 August 1 using multi-wavelength observations from Atmospheric Imaging Assembly (AIA), Chinese H$\alpha$ Solar Explorer (CHASE), and Interface Region Imaging Spectrograph (IRIS), combined with a self-consistent time-dependent magnetofrictional (TMF) model and magnetohydrodynamic (MHD) simulation. Spectral diagnostics reveal coexisting red and blue shifts along the edges and central axis of the jet, indicating helical plasma motion within a twisted magnetic structure. Numerical simulations demonstrate that the jet’s rotation arises from plasma propagating along helical open field lines, formed via reconnection between a pre-existing flux rope and overlying magnetic fields. Contrary to classical untwisting models, both linear and rotational velocities decrease with altitude during the jet propagation. These results highlight that the observed rotation results from plasma spiral motion along twisted fields rather than untwisting dynamics of the magnetic field itself, providing new insights into solar jet energetics and their connection to broader solar phenomena.
\end{abstract}


\section{Introduction}
Solar jets represent one of the most frequent eruptive phenomena in the solar atmosphere. Typically manifesting as collimated plasma ejections along linear trajectories (\citealt{Raouafi2016}; \citealt{Shen2021}), these events are fundamentally associated with magnetic reconnection processes (\citealt{Shibata1992}; \citealt{Priest1994}). Investigations reveal their critical role in transporting substantial energy from the chromosphere to the corona (\citealt{Brueckner1983}; \citealt{Shibata2007}; \citealt{Samanta2019}), making them important to study for understanding both magnetic reconnection dynamics and potential coronal heating mechanisms. Beyond their intrinsic significance, jets exhibit strong physical connections with diverse solar phenomena, such as plumes (\citealt{Lites1999}; \citealt{Raouafi2008}; \citealt{Raouafi2014}), high-energy particle events (\citealt{Aurass1994}; \citealt{Kundu1995}; \citealt{Li2011}; \citealt{Chen2013}; \citealt{Bucik2014}; \citealt{Vlahos2019}), solar wind (\citealt{Cirtain2007}; \citealt{McIntosh2011}; \citealt{McIntosh2012}; \citealt{Yu2016}), coronal mass ejections (\citealt{Gilbert2001}; \citealt{Jiang2008}; \citealt{Nistico2009}; \citealt{Panesar2016}; \citealt{Miao2018}), and Magnetohydrodynamic (MHD) waves (\citealt{Sarker2016}; \citealt{shen2018}; \citealt{Li2020}). This broad interconnectivity makes jet studies as an effective approach for investigating multiple aspects of solar activity.

Theoretical frameworks for jet formation have evolved significantly through seminal contributions. \cite{Shibata1992}, \cite{Priest1994}, and \cite{Yokoyama1995} established the foundational magnetic reconnection model, proposing that jets originate from reconnection-triggered energy release. This model was further developed by \cite{Canfield1996} who formulated the standard jet model, suggesting that jets result from reconnection between emerging flux ropes and external open fields. Crucially, when the initial flux rope contains twisted fields, reconnection transfers this twist to the open field configuration. The subsequent untwisting process produces observable rotational motions in jets, with the rotation direction opposing the initial twist, which is consistent with the observation. Model refinement continued with the blowout jet model of \cite{Moore2010}, extending the standard model by emphasizing the role of current sheets along flux rope surfaces, particularly the formation of curtain-like multi-stranded structures. Synthesis by \cite{Sterling2015} reconciles these perspectives, demonstrating how both standard and blowout jets can be unified within a single eruptive framework. Their analysis indicates that successful filament eruptions produce blowout jet characteristics, while confined eruptions correspond to standard jet features.

In addition to the vertical movement of the jet along the ejection direction, many studies have shown that some jets also have rotational motion (\citealt{Xu1984}; \citealt{Jibben2004}; \citealt{Shimojo2007}; \citealt{Liu2009}; \citealt{Shen2011}; \citealt{Zhang2014}; \citealt{Liu2014}). In addition, there are many structures on the Sun that also have torsional motion, such as spicules (\citealt{Pontieu2012}), tornadoes (\citealt{Wedemeyer2013}; \citealt{Su2014}), and surges (\citealt{Schmieder2013}; \citealt{Xue2016}). Current understanding attributes the rotational behavior to many possible mechanisms. The first and most widely accepted view, consistent with what we mentioned in the standard jet model above, is the rotation caused by the outer open field untwisting. Another possibility arises from numerical simulations of solar tornadoes (\citealt{Wedemeyer2012}), where twisted flux ropes reconnecting with outer fields create helical magnetic structures. In this scenario, plasma accelerated upward by magnetic pressure gradients and inertia would naturally follow spiral paths within the rotating magnetic structure. These mechanisms may coexist, with their relative contributions depending on the initial magnetic configuration and energy release conditions.

Data-constrained models and data-driven models, as numerical models using observational data as input, have proven effective in studying active region evolution and eruptive events (\citealt{Wu2006}; \citealt{Kliem2013}; \citealt{Amari2014}; \citealt{Nayak2019}; \citealt{Jiang2022}; \citealt{GuoY2024}). Data-constrained models typically employ non-linear force-free field (NLFFF) extrapolation methods to construct initial magnetic fields, then maintain fixed boundaries or provide numerical extrapolation through photospheric or lower coronal boundaries. In contrast, data-driven models utilize time series of magnetic fields, velocity fields, and/or electric fields in the photosphere as input data, directly driving the evolution of coronal magnetic fields and plasma through observational constraints. To investigate the formation and evolution of solar jets, data-driven models have also been widely applied in jet simulations (\citealt{Cheung2015}; \citealt{Jiang2016b}).

This study employs  multi-wavelength observations, spectroscopic diagnostics, and MHD simulations to investigate the rotational dynamics of solar jets. Our investigation aims to quantify the jet's rotational characteristics through spectral analysis of Doppler signatures, and identify the dominant rotation mechanisms by combining observation and numerical simulation. The case observations and simulation setup are introduced in Section \ref{section2} and Section \ref{section3}, respectively. Section \ref{section4.1} presents the spectral analysis, Section \ref{section4.2} show the results of simulation, and Section \ref{section4.3} quantifies the mechanical energy budget of the jet. The concluding Section \ref{section5} summarizes and discusses our results.

\section{Observations}\label{section2}
\begin{figure*}
   \centering
   \includegraphics[width=0.99\textwidth]{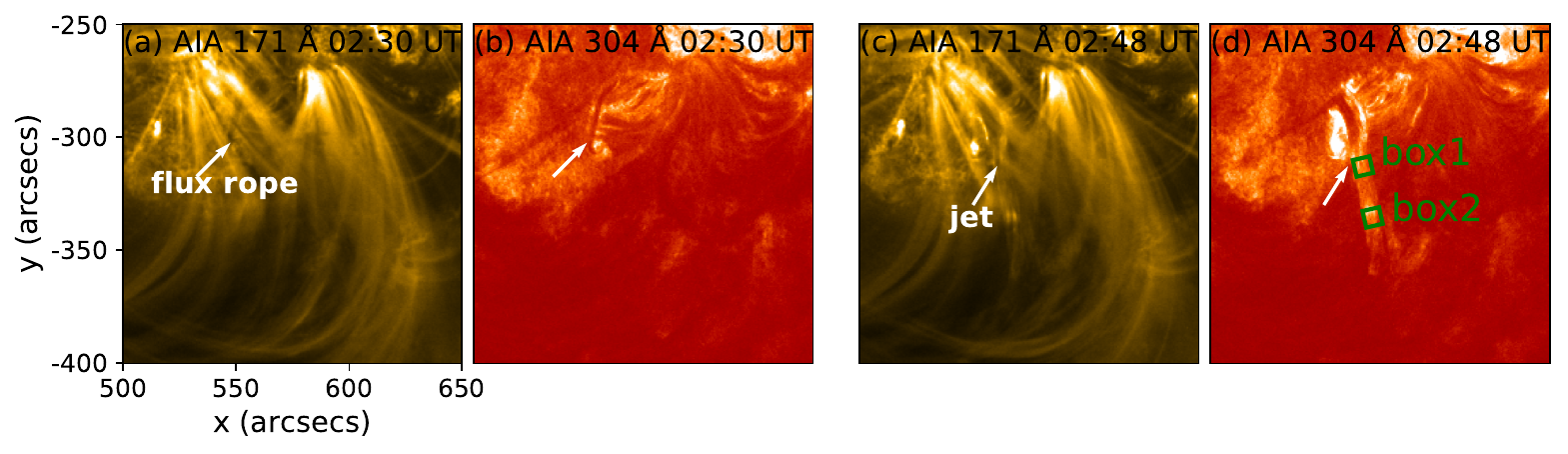}
   \caption{Jet formation process. (a)--(b) SDO/AIA observations in 171 $\mathrm{\AA}$ and 304 $\mathrm{\AA}$ at 02:30 UT before the jet formation; (c)--(d) Corresponding 171 $\mathrm{\AA}$ and 304 $\mathrm{\AA}$ observations during the jet eruption at 02:48 UT. The attached animation, displaying AIA 171 $\mathrm{\AA}$ and 304 $\mathrm{\AA}$ images from 02:30 UT to 03:15 UT, clearly reveals the entire sequence of events: the presence and subsequent uplift of a small filament at the base prior to the eruption, followed by its eruption, the triggering of the jet, and the rotational motion of the jet.}
    \label{fig1}%
\end{figure*}

We analyze a rotating jet event occurring in Active Region 13380 around 02:40 UT on 2023 August 1, and we find that there is no corresponding CME after checking observations from the white light coronagraph aboard the Solar TErrestrial RElations Observatory (STEREO; \citealt{Kaiser2008}) and Large Angle and Spectroscopic Coronagraph (LASCO) instrument (\citealt{Brueckner1995}) aboard Solar and Heliospheric Observatory (SOHO). This event is observed simultaneously by Solar Dynamics Observatory (SDO; \citealt{Pesnell2012}), Chinese H$\alpha$ Solar Explorer (CHASE; \citealt{Li2022}), and  Interface Region Imaging Spectrograph (IRIS; \citealt{Pontieu2014}). The multi-wavelength observations of Atmospheric Imaging Assembly (AIA; \citealt{Lemen2012}) aboard SDO reveal its clockwise rotational motion (see Figure \ref{fig1} animation). Figures \ref{fig1}(a)--(b) display a magnetic flux rope at the bottom before the jet occurs, and Figures \ref{fig1}(c)--(d) document the jet eruption. CHASE provides full-disk spectral observations of both the Fe I line and H$\mathrm{\alpha}$ line (\citealt{Li2022}; \citealt{Qiu2022}). In this study, we primarily use the H$\mathrm{\alpha}$ spectral data to analyze chromospheric dynamics of the jet. Complementary observations from IRIS include high-resolution images and Si IV spectral line measurements of the same jet, which enable us to determine the jet's velocity in the transition region.

\begin{figure*}
   \centering
   \includegraphics[width=0.8\textwidth]{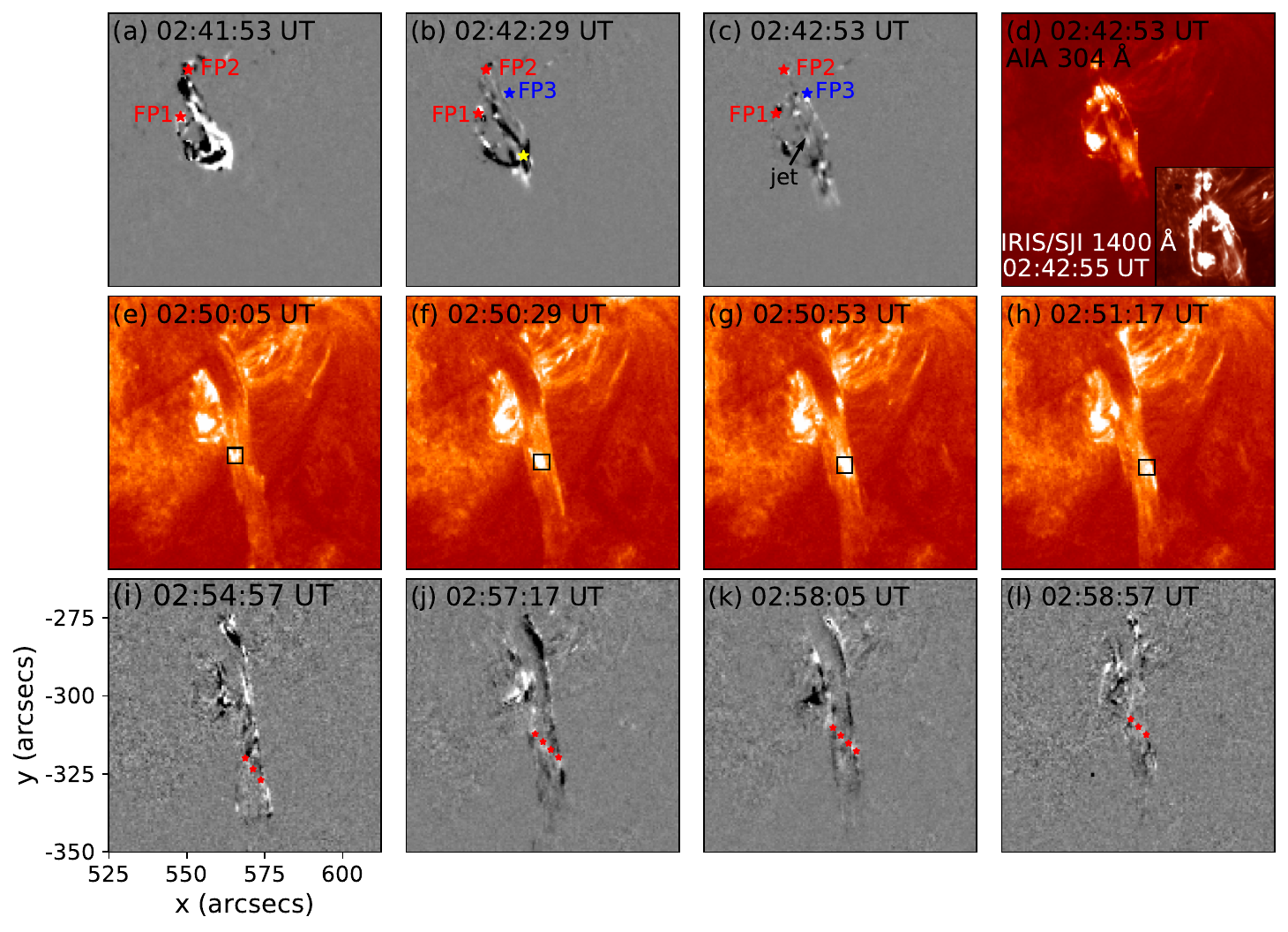}
   \caption{Jet dynamics in multi-wavelength observations. (a)--(c) Running difference AIA 304 $\mathrm{\AA}$ images showing pre-jet configuration: FP1 and FP2 denote twin footpoints of a twist magnetic flux rope, and FP3 marks one footpoint of the external open field lines, with yellow stars indicating magnetic reconnection sites. (d) coordinated observations of AIA 304 $\mathrm{\AA}$ and IRIS/SJI 1400 $\mathrm{\AA}$. (e)--(h) Jet eruption phase of AIA 304 $\mathrm{\AA}$ images with black rectangles highlighting bright plasma blocks. (i)--(l) Running difference AIA 304 $\mathrm{\AA}$ images during jet descending phase, with red tracers follow plasma motion trajectories.}
    \label{fig2}%
\end{figure*}

We categorize the jet evolution into three distinct phases: generation, ascent, and descent, as illustrated in Figure \ref{fig2}. Figures \ref{fig2}(a)--(d) show the generation phase of the jet. At 02:41 UT, the magnetic flux rope begins to rise. Figure \ref{fig2}(a) displays a rising magnetic flux rope, where FP1 and FP2 denote its two footpoints. In Figure \ref{fig2}(b), this small flux rope undergoes magnetic reconnection with external open fields. The blue marker FP3 indicates one footpoint of the initial open field line (with the other located outside the field of view), while the yellow marker shows the reconnection site. It should be noted that the term open field does not refer to a truly open magnetic structure in the global sense. It describes magnetic field lines that are open within the current field of view. These field lines may indeed be closed globally, but their other footpoints lie outside the simulated or observed domain. For collimated jets, their outer spine is sufficiently extended that it can be treated as effectively open (\citealt{Shen2021}).
Figure \ref{fig2}(c) reveals that after reconnection, the original flux rope footpoint FP1 and the original open field footpoint FP3 reconnect to form a bright loop, while a jet propagates along the newly formed open field line with its footpoint FP2. Both AIA 304 $\mathrm{\AA}$ and IRIS slit-jaw imager (SJI) 1400 $\mathrm{\AA}$ observations in Figure \ref{fig2}(d) clearly show the jet and the newly formed loop structure at the jet base.

Following the jet formation, Figures \ref{fig2}(e)--(h) track the subsequent plasma motion. Time-series images reveal a bright plasma block (demarcated by black rectangles) propagating from upper-left to lower-right along a left-hand helical trajectory. Next, we show the images of the jet descending phase (refer to Figure \ref{fig9}(b) to confirm the ascending and descending stages of the jet) in Figures \ref{fig2}(i)--(l). By examining the animation corresponding to Figure 1, it can be observed that plasma fallback begins around 02:52 UT, which is generally consistent with the fallback time shown in Figure \ref{fig9}(b). As for this descending motion, it can be explained by plasma moving upward along the magnetic field lines to its maximum height, decelerating, and then beginning to fall back under the influence of gravity. Studies such as \cite{Robustini2016}, \cite{Raouafi2016} and \cite{Miao2018} have indicated that such fallback is a common phenomenon in jets.
While the descending matter is sparse, we can still find that some residual plasma in the jet moves across the central axis from the bottom right to the top left (indicated by the red dots). We propose that sustained twisting of open field lines throughout the whole process of the jet governs this bidirectional motion: plasma ascends/descends along the helical field lines during respective phases. 

\section{Modeling Description}\label{section3}
To simulate the jet formation process and investigate its rotation mechanism through numerical simulation, we emphasize that the bottom magnetic flux rope should develop self-consistently in our simulation, which differs from methods that artificially insert a pre-existing magnetic flux rope and subsequently induce its eruption. We therefore employ the time-dependent magnetofrictional (TMF) method integrated with the data-driven thermodynamic MHD model for our simulation framework (detailed implementation can be found in \citealt{Guo2024}).

First, we employ the TMF model to investigate the long-term evolution of Active Region 13380. This aims to generate the magnetic flux rope self-consistently through the natural evolution of the active region, avoiding artificial insertion methods. The initial magnetic field is the potential field extrapolated by Message Passing Interface Adaptive Mesh Refinement Versatile Advection Code \footnote{\url{https://amrvac.org/}} (MPI-AMRVAC; \citealt{Keppens2012}, \citealt{Porth2014}, \citealt{Xia2018}, \citealt{Keppens2023}). The computational domain spans [$x_{min},x_{max}$] $\times$ [$y_{min},y_{max}$] $\times$ [$z_{min},z_{max}$] = [-47.6,47.6] $\times$ [-47.6,47.6] $\times$ [1,95.2] Mm$^3$ with $260 \times 260 \times 260$ uniform grids, maintaining the Helioseismic and Magnetic Imager (HMI) spatial resolution of 0.5$^{\prime\prime}$ per pixel. The bottom boundary condition of the potential field is specified using the vertical magnetic field component $B_z$ from the HMI photospheric vector magnetogram observed at 00:00 UT on 2023 July 31.

Subsequently, we utilize the hmi.sharp\_cea\_720s magnetic map covering 2023 July 31 00:00 UT to 2023 August 1 02:24 UT as the driving boundary condition at the photosphere. The photospheric velocity field is derived using the Differential Affine Velocity Estimator for Vector Magnetograms (DAVE4VM; \citealt{Schuck2008}) method. These vector magnetic fields and velocity fields are directly imposed on the inner ghost layer, while the outer ghost layer values are determined via a zero-gradient extrapolation. The governing equations for the TMF stage can be expressed as:

\begin{equation}
\begin{gathered}
\begin{split}
\frac{\partial\boldsymbol{B}}{\partial t}+\nabla\cdot(\boldsymbol{v}\boldsymbol{B}-\boldsymbol{B}\boldsymbol{v})=-\nabla\times(\eta\boldsymbol{J}), 
\\ \boldsymbol{v}=\frac{1}{\nu}\frac{\boldsymbol{J}\times\boldsymbol{B}}{B^{2}},
\\ \nu=\frac{\nu_0}{1-e^{-z/L}},  
\end{split}
\end{gathered}
\end{equation}
where $\boldsymbol{B}$ is the magnetic field strength, $\boldsymbol{v}$ is the velocity, and $\boldsymbol{J}$ is the electric current density. The viscous coefficient of the friction $\nu_0=10^{-15}$ s cm$^{-2}$, the viscous decay length $L=10$ Mm, and the magnetic diffusivity $\eta$ (\citealt{Cheung2012}; \citealt{Pomoell2019}) are employed. The implemented anomalous resistivity are shown as follows:

\begin{equation}
\eta=\eta_0+\eta_1\frac{50\zeta}{1+e^{-2(\zeta-3.0)}}.
\end{equation}
We employ two distinct diffusion coefficients: $\eta_0=2\times10^{11}$ cm$^2$ s$^{-1}$ serves to improve computational robustness, while  $\eta_1=2\times10^{12}$ cm$^2$ s$^{-1}$ addresses the diffusion in current sheets characterized by significant $J/B$. The latter coefficient incorporates a scaling factor $\zeta=J^2B^{-2}\triangle^2$, where $\triangle=\min[\triangle x,\triangle y,\triangle z]$ denotes the minimum spatial grid spacing across orthogonal coordinate axes.

The eruptive dynamic is further simulated by a data-driven magnetohydrodynamic model developed in recent studies (\citealt{Guo2023}; \citealt{Guo2024}). The model initialization utilizes magnetic field inherited from the TMF model captured at 02:24 UT on 2023 August 1. Throughout the computational evolution, data-driven boundary conditions are adopted (\citealt{Guo2019}). A thermodynamic MHD model (\citealt{Xia2016}) is employed to simulate the dynamic process, which is governed by the following equations:

\begin{equation}
\begin{gathered}
\begin{split}
\frac{\partial\rho}{\partial t}+\nabla\cdot(\rho \boldsymbol{v})=0, \\
\frac{\partial(\rho\boldsymbol{v})}{\partial t}+\nabla\cdot\left(\rho\boldsymbol{v}\boldsymbol{v}+p_{\mathrm{tot}}\boldsymbol{I}-\frac{\boldsymbol{B}\boldsymbol{B}}{\mu_{0}}\right)=\rho\boldsymbol{g}, \\
\frac{\partial \boldsymbol{B}}{\partial t}+\nabla\cdot(\boldsymbol{vB}-\boldsymbol{Bv})=0, \\
\frac{\partial\varepsilon}{\partial t}+\nabla\cdot\left(\varepsilon\boldsymbol{v}+p_{\mathrm{tot}}\boldsymbol{v}-\frac{\boldsymbol{BB}}{\mu_0}\cdot\boldsymbol{v}\right) \\
=\rho \boldsymbol{g}\cdot\boldsymbol{v}+H_0e^{-z/\lambda}-n_\mathrm{e}n_H\Lambda(T) \\
+\nabla\cdot(\boldsymbol{\kappa}\cdot\nabla T).
\end{split}
\end{gathered}
\end{equation}
The critical components of the governing equation are as follows. The total pressure $p_\mathrm{tot}\equiv p+B^2/(2\mu_0)$ is formulated under the fully ionized plasma assumption. The gravitational acceleration $\boldsymbol{g}=-g_\odot r_\odot^2/(r_\odot+z)^2\boldsymbol{e_z}$, where $g_\odot=274$ m $\mathrm{s}^{-2}$ represents solar surface gravity and $r_{\odot}$ denotes the solar radius. $\varepsilon=\rho{v}^{2}/2+p/(\gamma-1)+B^{2}/(2\mu_{0})$ is the total energy density. The thermal conduction governed by the field-aligned thermal conduction conductivity $\boldsymbol{\kappa}=\kappa_\parallel\hat{\boldsymbol{b}}\hat{\boldsymbol{b}}$ and the Spitzer heat conductivity $\kappa_\parallel=10^{-6}$ T$^{\frac{5}{2}}$ erg cm$^{-1}$ s$^{-1}$ K$^{-1}$. Energy exchange mechanisms comprise optically thin radiative losses $n_{e}n_{H}\Lambda(T)$ and an  empirical heating $H_0e^{-z/\lambda}$ that sustains coronal thermal equilibrium, where $H_0=10^{-4}$ erg cm$^{-3}$ s$^{-1}$, $\lambda=60$ Mm.

The initial atmosphere employs a hydrostatic atmospheric model from the chromosphere to the corona, which is shown as follows:
\begin{equation}
    \begin{aligned}
    T\left(z\right)=&\begin{cases}T_{\mathrm{ch}}+\frac{1}{2}(T_{\mathrm{co}}-T_{\mathrm{ch}})(\tanh\left(\frac{z-h_{\mathrm{tr}}-0.27}{w_{\mathrm{tr}}}\right)+1)&z\leqslant h_{tr},\\\\\left(\frac{7}{2}\frac{F_{\mathrm{c}}}{\kappa}(z-h_{\mathrm{tr}})+T_{\mathrm{tr}}^{7/2}\right)^{2/7}&z>h_{tr}.\end{cases}
    \end{aligned}
\end{equation}
The atmospheric initialization specifies these critical parameters: chromospheric temperature $T_{\mathrm{ch}}=8000$ K, coronal temperature  $T_{\mathrm{co}}=1.5$ MK along with transition region height $h_{\mathrm{tr}}=2$ Mm and thickness $w_{\mathrm{tr}}=0.2$ Mm. The thermal conduction flux is prescribed as $F_{\mathrm{c}}=2\times10^{5}$ erg cm$^{-2}$ s$^{-1}$. Density distribution derives from the bottom number density $1.15\times10^{15}$ cm$^{-3}$. Following previous works (\citealt{Jiang2016a}; \citealt{Kaneko2021}; \citealt{Guo2023}), we reduce the observed magnetic field strength by a factor of 10 to optimize computational performance and reduce numerical dissipation.

\section{Results}
\subsection{Spectral Analysis}\label{section4.1}

\begin{figure*}
   \centering
   \includegraphics[width=0.99\textwidth]{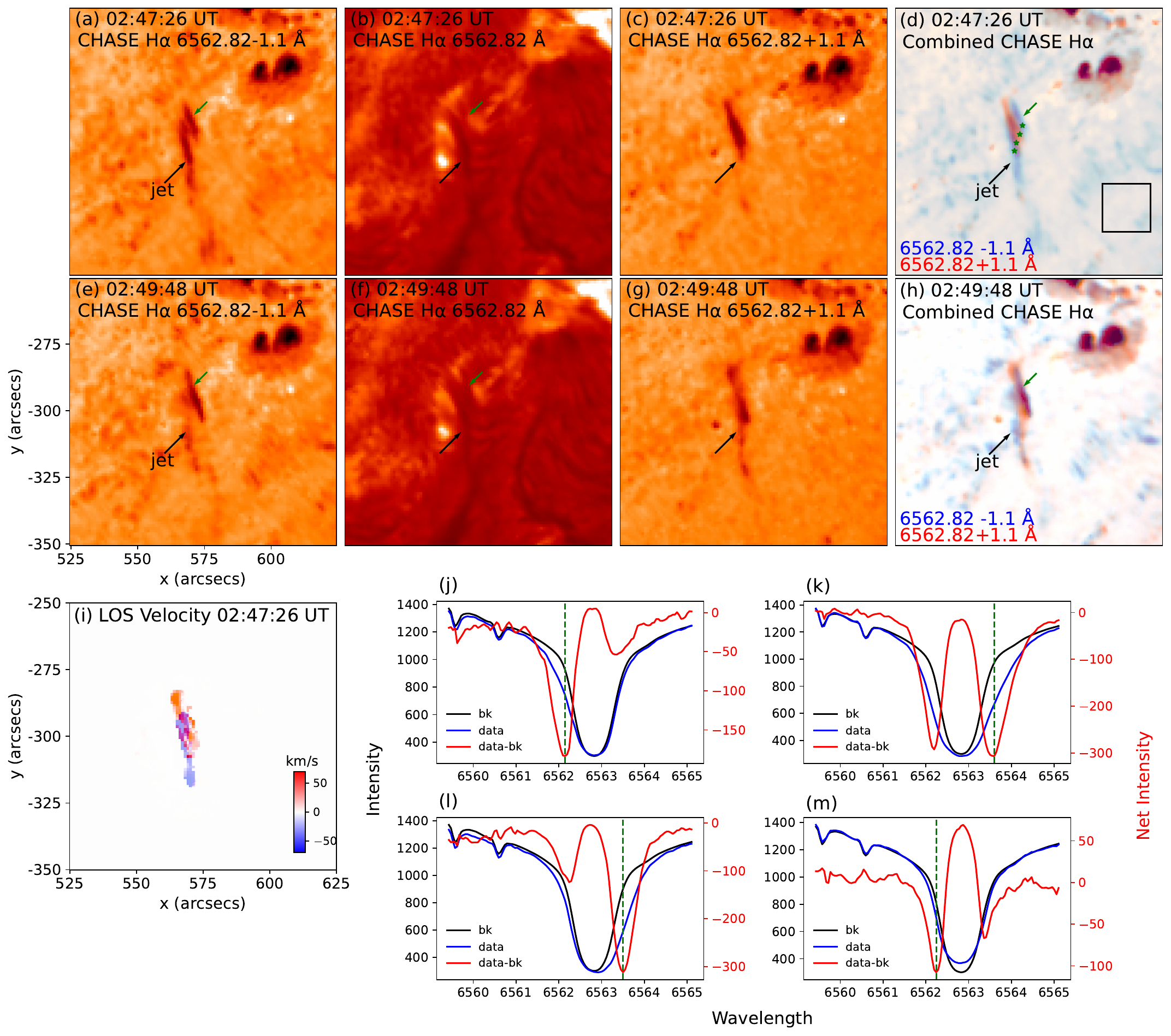}
   \caption{CHASE H$\alpha$ spectroscopic diagnostics of the jet dynamics. (a)--(d) CHASE H$\alpha$ blue-wing (6562.82-1.1 Å), core (6562.82 Å), red-wing (6562.82+1.1 Å), and composite blue-red wing images at 02:47:26 UT. The black box shows the background profile reference region. (e)--(h) Corresponding spectral components at 02:49:48 UT. The black arrow marks the position of the jet, the green arrow marks a blue shift region to the right of the jet. (i) Doppler velocity map at 02:47:26 UT. (j)--(m) The original spectral line profiles of the four points from left to right in panel (d) and the net spectral line profiles after subtracting the background line profile. The green dashed line marks the wavelength corresponding the minimum intensity of the net spectral line profile.}
    \label{fig3}%
\end{figure*}

We show the monochromatic images of CHASE H$\mathrm{\alpha}$ blue wing ($6562.8-1.1$ $\mathrm{\AA}$), line core (6562.8 $\mathrm{\AA}$), and red wing ($6562.8+1.1$ $\mathrm{\AA}$) at two moments in Figure \ref{fig3}, which show that jet appears in both red wing and blue wing simultaneously. Comparative analysis between Figures \ref{fig3}(a) and (c) at 02:47 UT and Figures \ref{fig3}(e) and (g) at 02:49 UT reveals dynamic spectral asymmetry: blue-wing dominance at 02:47 UT switching to red-wing predominance two minutes later, indicating that plasma in the jet flow is constantly moving in the blue wing and the red wing. To investigate spatial correlations, we overlay the diagrams of the blue wing and red wing in Figures \ref{fig3}(d) and \ref{fig3}(h). The composite maps show partial overlap along the central axis, and distinct left-right segregation with blue-wing dominating the left side and red-wing on the right side.

Figures \ref{fig3}(j)--(m) display original and background-subtracted spectral line profiles from the four green points marked in Figure \ref{fig3}(d). The background is the average spectral line profile of the black box area shown in Figure \ref{fig3}(d). As seen in Figure \ref{fig3}(j), the blue wing of the spectral line on the left side of the jet exhibits strong absorption, while Figure \ref{fig3}(l) reveals prominent red wing absorption on the right side.
Figure \ref{fig3}(k) demonstrates the central region of the jet, displays significant line broadening. It is important to note that this broadening is unlikely to arise from microscopic turbulence, as turbulent broadening effects would be expected to remain relatively uniform at adjacent locations within the same magnetic flux rope. The observed transition from blueshift dominance on the left side of the jet to redshift dominance on the right, with the central region exhibiting spectral features attributable to both components, indicates that the broadening likely results from large-scale bulk motions rather than localized turbulent effects. 

Figure \ref{fig3}(i) shows the Doppler velocity diagram of the jet at 02:47:26 UT. The method used to calculate the Doppler velocity is to find the wavelength $\lambda_i$ corresponding to the minimum intensity of the net spectral line profile, and the Doppler velocity is calculated through $v$$\mathrm{=\frac{c({\lambda }_{i}-{\lambda }_{0})}{{\lambda }_{0}}}$. The reference wavelength $\lambda_0$ is the line center calculated by using the 80\% bisector method (see details in \citealt{Zhao2022}) on the background line profile. By overlaying velocities derived from blue-wing and red-wing minima in Figure \ref{fig3}(i), we visualize the red/blue shifts dominate on the jet's right/left sides, respectively, with the red and blue shift in the central-axis simultaneously. This corresponds to the observed clockwise rotation of the jet and its complex magnetic topology.

Figures \ref{fig3}(a) and \ref{fig3}(e) reveal an additional blue-wing absorption component (green arrow) on the right side of the jet, with its corresponding spectral profile shown in Figure \ref{fig3}(m). Figures \ref{fig3}(a)--(b) demonstrate this component's spatial distinction from the jet, the black arrow specifically marks the jet structure, whereas the green arrow indicates separate right-side absorption. The spectral characteristics in Figure \ref{fig3}(m) differ markedly from those in Figures \ref{fig3}(i)--(l). The original spectral line of this region has no obvious absorption or broadening, but the overall spectral line has an emission in the line center, resulting in absorption in the red/blue-wing after abstracting the background. We therefore identify this blue-wing material not as jet plasma, but may be a counterpart of chromosphere evaporation or reconnection outflow.

\begin{figure*}
   \centering
   \includegraphics[width=0.99\textwidth]{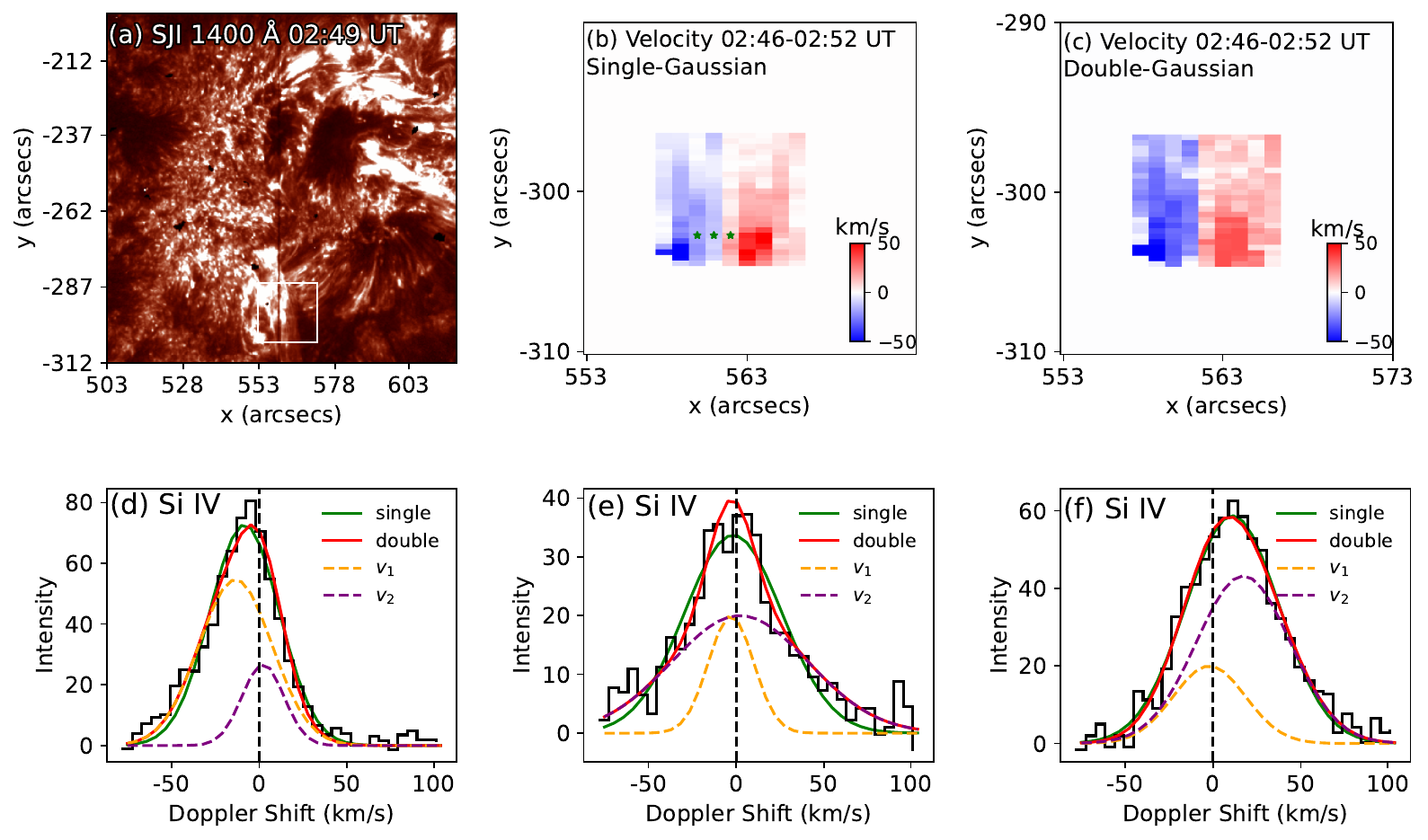}
   \caption{IRIS Si IV spectroscopic diagnostics of the jet dynamics. (a) Image of IRIS/SJI 1400 $\mathrm{\AA}$ at 02:49 UT. The attached animation, displaying IRIS SJI 1400 $\mathrm{\AA}$ observations from 02:30 UT to 03:15 UT, clearly reveals the filament prior to its eruption and the subsequent rotational motion of the jet following the eruption. (b)--(c) Doppler velocity maps within the jet in the white box in panel (a) calculated by single Gaussian fitting and double Gaussian fitting, respectively. (d)--(f) Spectral line profiles of the three points from left to right in panel (b). The observed profiles (black) with single-Gaussian (green) and double-Gaussian (red) fits, where yellow/purple dashed curves represent two components of the double Gaussian fit.}
    \label{fig4}%
\end{figure*}

Figure \ref{fig4}(a) presents the IRIS/SJI 1400 $\mathrm{\AA}$ observations of the jet, and the animation confirms consistent clockwise rotation with SDO/AIA observations in Figure \ref{fig1}. The spectral line profiles of the three points of the jet from left to right (labeled in Figure \ref{fig4}(b)) are shown in Figures \ref{fig4}(d)--(f). We use Doppler velocity instead of wavelength in the horizontal coordinate, and the Doppler velocity is calculated based on $v$$\mathrm{=\frac{c({\lambda }_{i}-{\lambda }_{0})}{{\lambda }_{0}}}$. We adopt the theoretical line center 1402.77 $\mathrm{\AA}$, because the result of Gaussian fitting of quiet-region mean spectral line profile shows agreement with theoretical values. The spectral line profile in Figure \ref{fig4}(e) shows two peaks, which lead us to use both single Gaussian fitting and double Gaussian fitting approaches, with resultant Doppler velocity maps shown in Figures \ref{fig4}(b) and \ref{fig4}(c), respectively. The $y$-axis resolution is 0.33$^{\prime\prime}$ and the $x$-axis resolution is 1.00$^{\prime\prime}$.  For double Gaussian results, we display velocities corresponding to the strongest spectral component. Both fitting methods consistently reveal red shift on the right side and the blue shift on the left side of the jet, which are consistent with the CHASE observation.

\subsection{Simulation Results}\label{section4.2}
\begin{figure*}
   \centering
   \includegraphics[width=0.7\textwidth]{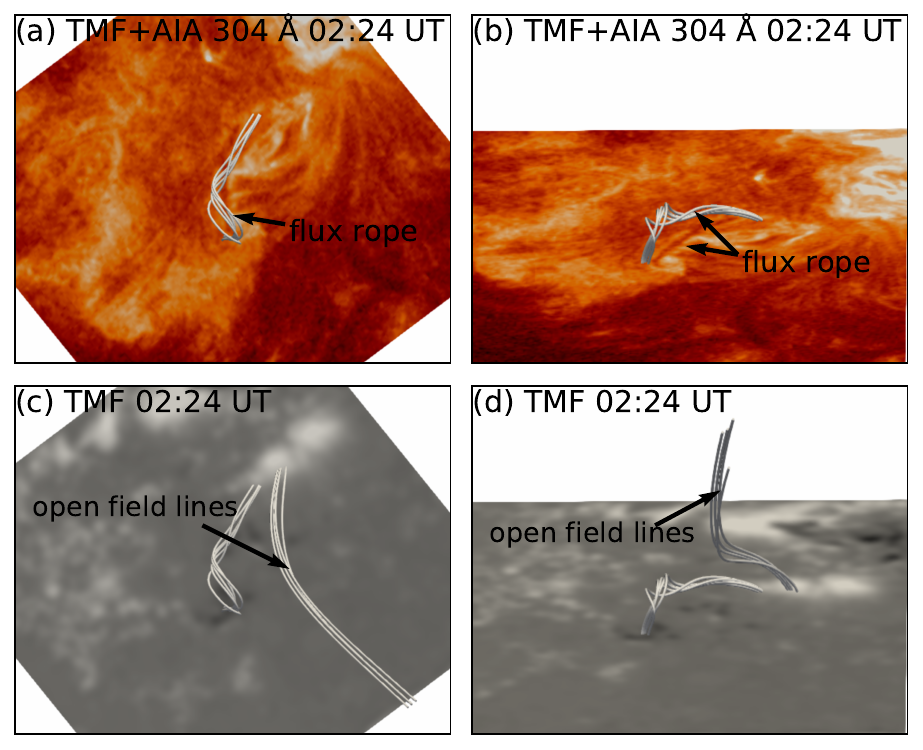}
   \caption{Magnetic Topology in the TMF model. (a)--(b) Magnetic flux rope morphology in TMF model superimposed on pre-eruption SDO/AIA 304 $\mathrm{\AA}$ background at 02:24 UT. (c)--(d) Open field lines in TMF model overlaid on HMI magnetogram at 02:24 UT. The arrows mark the magnetic flux rope and the open field lines.}
    \label{fig5}%
\end{figure*}

Figure \ref{fig5} presents the TMF simulation results at 02:24 UT on August 1, demonstrating the self-consistent formation of a twisted magnetic flux rope through more than 24 hours of evolution. We overlay it on AIA 304 $\mathrm{\AA}$ image at 02:24 UT, which shows spatial correspondence between the magnetic flux rope and the observed filament in Figures \ref{fig5}(a)--(b). Following the method of \cite{Guo2017} and \cite{Guo2021}, we calculated the twist number of 100 selected field lines relative to the central axis of the flux rope. By averaging their values and computing the standard deviation, we obtained a mean twist number of 1.40 $\pm$ 0.37 for the flux rope. This result falls within the typical range of twist numbers (1–3) observed in active regions.
Figures \ref{fig5}(c)--(d) display open magnetic field lines superimposed on HMI magnetogram at 02:24 UT. The open field lines are straight at this time.

\begin{figure*}
   \centering
   \includegraphics[width=0.8\textwidth]{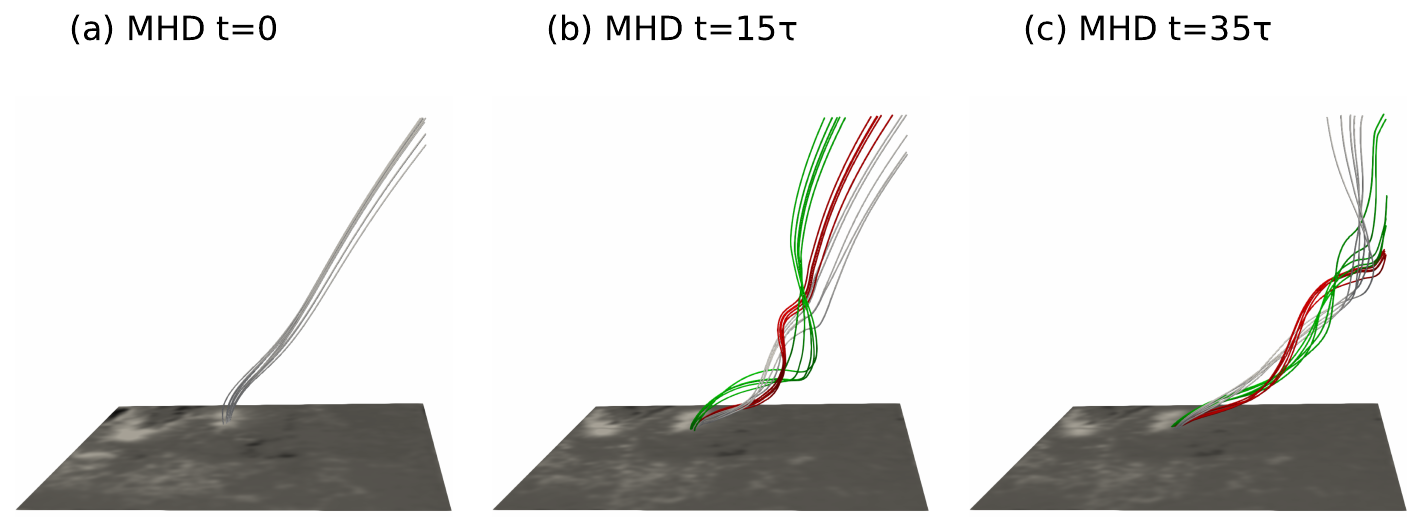}
   \caption{Magnetic topology evolution in the MHD model showing open field line evolution. (a) Initial configuration at $t=0$, (b) Twist transmission phase at $t=15$ $\tau$, (c) Final twisted configuration of the magnetic field at $t=35$ $\tau$. $\tau$ corresponds to 85 s in physical time.}
    \label{fig6}%
\end{figure*}

\begin{figure*}
   \centering
   \includegraphics[width=0.99\textwidth]{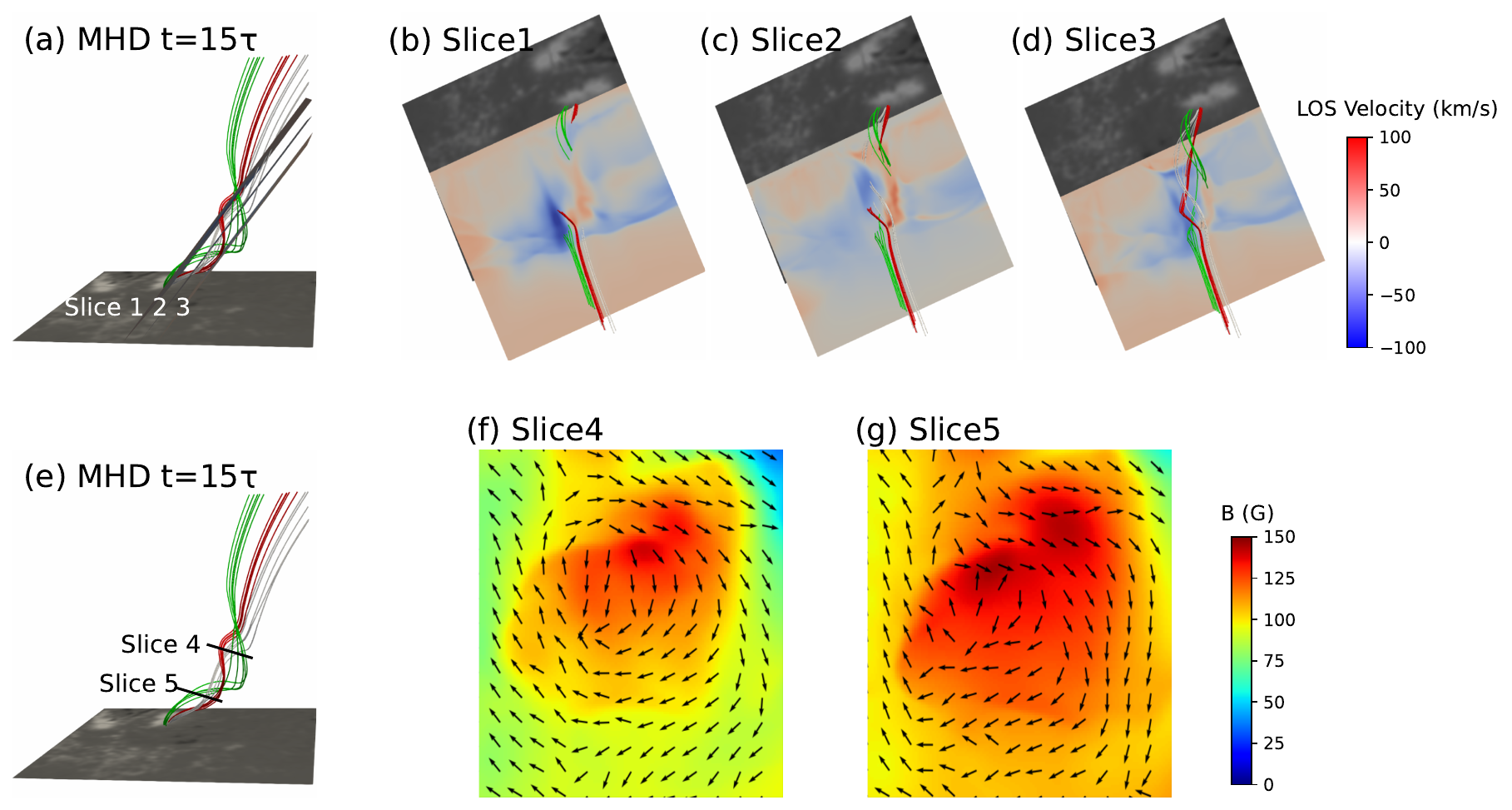}
   \caption{Plasma kinematics in the MHD Simulation. (b)--(d) The perpendicular components of velocity to the three slices in panel (a). (f)--(g) Plane velocities on the two slices in panel (e). The background is the magnetic field strength. }
    \label{fig7}%
\end{figure*}

\begin{figure*}
   \centering
   \includegraphics[width=0.99\textwidth]{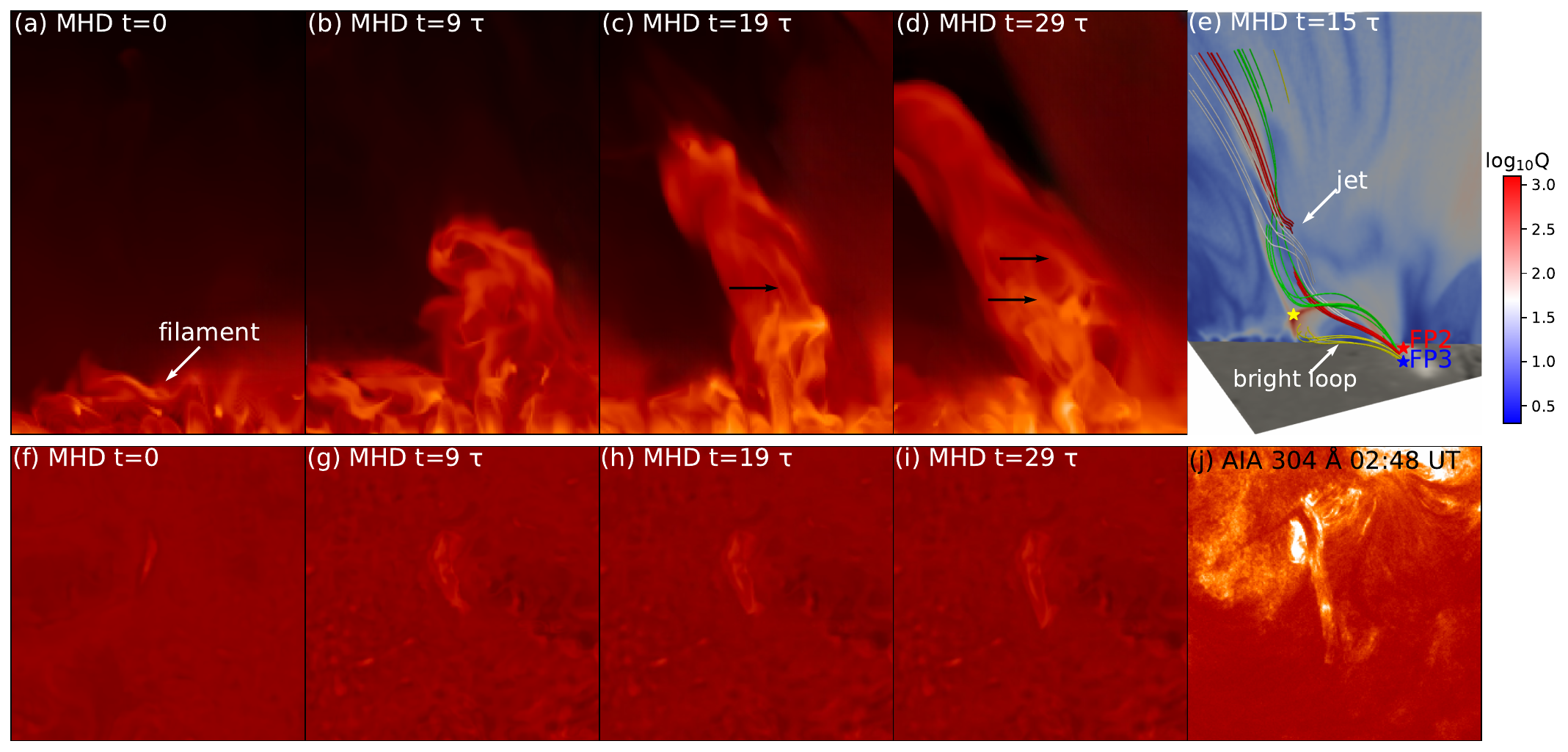}
   \caption{Synthetic jet images in 304 $\mathrm{\AA}$. (a)--(d) Side-view synthesized emission images at t=0, 9, 19, 29 $\tau$, black arrows track the movement of the bright blocks in the jet. The time t=0 corresponds to 02:24 UT, with $\tau$=85 s. The attached animation presents a side view of the synthetic emission images from t=0 to t=33 $\tau$, illustrating the complete process of jet formation due to the filament eruption. The rotational motion of bright features within the jet can be observed in the sequence. (e) Magnetic topology at MHD t=15 $\tau$ with squashing factor $Q$ slice, where jet, flux rope footpoint FP2 (red), open-field footpoint FP3 (blue), and bright loop (yellow) are annotated. (f)--(i) Corresponding top-view radiative synthesized images. (j) Observed AIA 304 $\mathrm{\AA}$ at 02:48 UT for comparison. The attached animation displays the evolution of the synthetic 304 $\mathrm{\AA}$ jet emission, showing the eruption of a solar filament and the subsequent formation of the jet. The rotational motion of the plasma within the jet can be clearly observed.}%
    \label{fig8}
\end{figure*}

We input the TMF model at 02:24 UT into the thermodynamic MHD model, which successfully induces a rapid magnetic flux rope eruption. Figure \ref{fig6} shows the evolution of open magnetic field lines during this process. Figures \ref{fig6}(a) and \ref{fig6}(b) reveal a gradual transformation from straight open field lines to twisted ones. This evolution corresponds to the underlying physical process where the ascending flux rope reconnects with overlying open fields, thereby transferring twist to the open field lines. In our simulation, the flux rope ascends and undergoes reconnection with the open field within approximately 5 minutes.
Figure \ref{fig6}(c) shows that the twist is transmitted to the entire open field lines. Following the magnetic frozen-in principle, jet plasma subsequently travels along these twisted field lines, producing the observed rotational kinematics. Crucially, the simulated left-handed helical configuration of the flux rope directly corresponds to observed left-handed rotational motion in the jet, establishing consistency between numerical simulation and observation. We also computed the temporal evolution of the magnetic twist in these open fields penetrating the jet, following the method of \cite{Guo2017} and \cite{Guo2021}. The results confirm that the open fields exhibited no significant twist reduction during the entire jet process, indicating the absence of field untwisting. 

Figure \ref{fig7} presents the jet velocity in the MHD simulations. Three cross-sectional slices aligned with the observational view angle are created at distinct altitudes as shown in Figure \ref{fig7}(a). Figures \ref{fig7}(b)--(d) display the simulated velocity component perpendicular to each respective slice. Assuming we are viewing the slice plane vertically, this velocity component corresponds to the simulated velocity along our current line of sight.
The velocity distributions demonstrate redshift on the jet's right side and blueshift on the left side in Figures \ref{fig7}(b)--(d), matching the line-of-sight velocity patterns observed by CHASE and IRIS in Section \ref{section4.1}. Crucially, these Doppler shifts exhibit altitude-dependent variations, explaining the observed bidirectional velocity despite the inherent complexity of spectral line formation. This behavior results from the twisted magnetic topology rather than a simple cylindrical magnetic flux tube, where the twisted field lines govern plasma motion. In Figure \ref{fig7}(e), we select two slices perpendicular to the jet, and show their in-plane velocity vector fields superimposed on magnetic field intensity maps in Figures \ref{fig7}(f) and \ref{fig7}(g). The velocity pattern shows clockwise rotational motion at different projected heights within the helical magnetic structure viewing against the jet propagation direction, directly matching the left-handed helicity of the magnetic flux rope.

To compare our simulation results with observations, we conduct radiation synthesis using Radiation Synthesis Tools\footnote{\url{https://github.com/fuwentai/radsyn_tools}}. Figure \ref{fig8} presents the synthesized 304 $\mathrm{\AA}$ optical thick images. The side view evolution of the jet is displayed in Figures \ref{fig8}(a)--(d), where we observe the rising magnetic flux rope at the base that eventually forms the jet. Notably, some bright block movement within the jet (indicated by black arrows) follows a left-hand spiral pattern matching observed material motion. The supplementary animation of Figure \ref{fig8} reveals the process more completely, showing multiple distinct plasma blobs undergoing rotational motion within the jet. Their direction of rotation is consistent with the observed rotational direction.
Figure \ref{fig8}(e) illustrates the magnetic configuration and squashing factor $Q$ calculated by CuQSL\footnote{\url{https://github.com/gychen-NJU/CuQSL}} at $t=15$ $\tau$. This reveals both the open field lines and underlying magnetic flux rope, corresponding to the bright loop structure beneath the jet in Figure \ref{fig2}(d). Key features including the null point, jet footpoint FP2, and bright loop footpoint FP3 are annotated for comparison with Figure \ref{fig2}. The synthesized 304 $\mathrm{\AA}$ images from the observational perspective shown in Figures \ref{fig8}(f)--(i) demonstrate reasonable agreement with the AIA 304 $\mathrm{\AA}$ observation shown in Figure \ref{fig8}(j), confirming the consistency between our simulation and actual observations.

\subsection{Mechanical Energy Budget}
\label{section4.3}

\begin{figure*}
   \centering
   \includegraphics[width=0.99\textwidth]{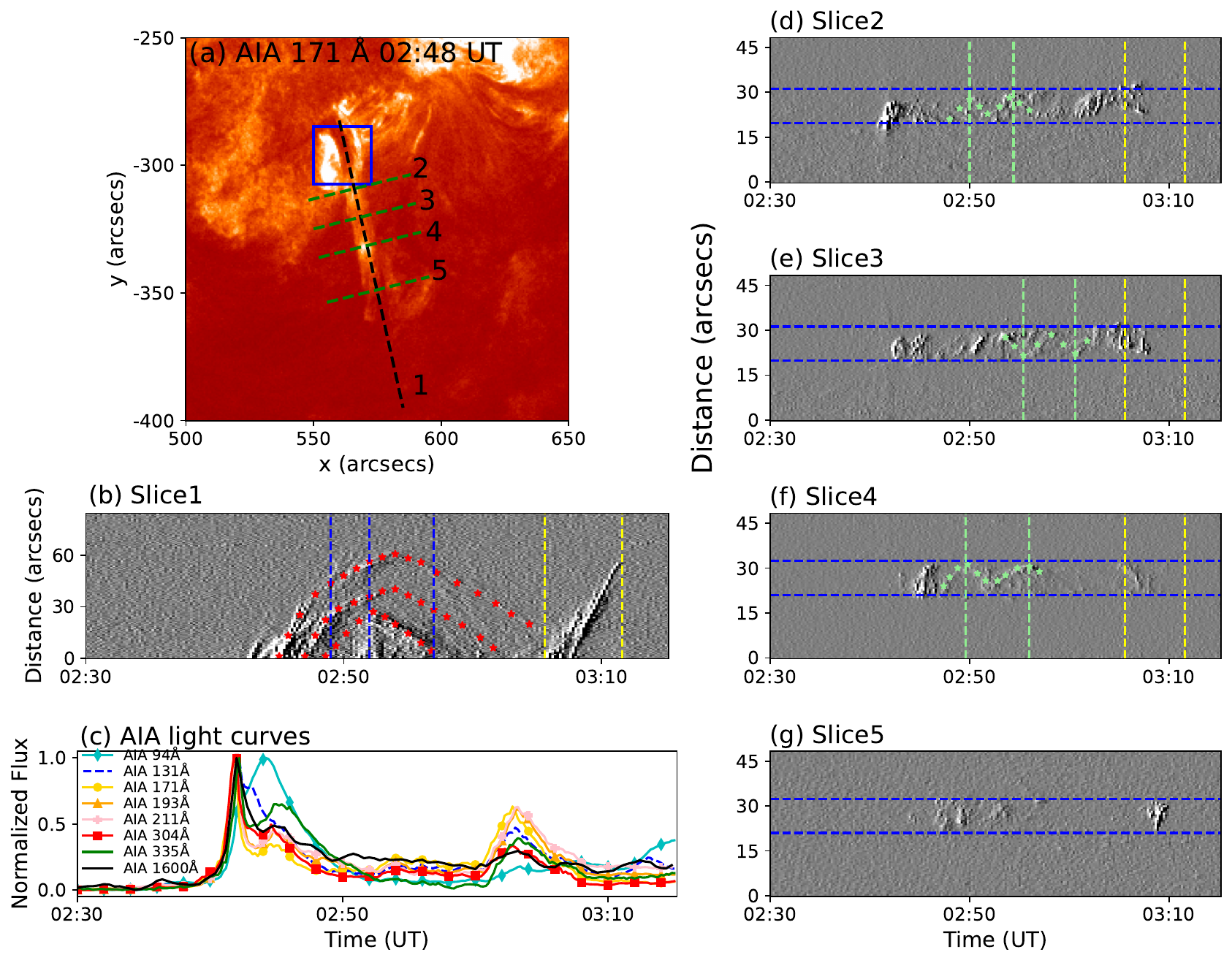}
   \caption{Jet dynamic analysis. (a) SDO/AIA 304 $\mathrm{\AA}$ image at 02:48 UT with black dashed lines indicating the position of the axial slit and four green dashed lines indicating the four slits perpendicular to the jet axis. The blue box indicates the region used for calculating the AIA light curves in panel (c). (b) Axial time-distance diagram (Slice 1). The red dots represent the three jet thread trajectories, and the three blue vertical lines represent the three moments at 02:49 UT, 02:52 UT, 02:57 UT. The two yellow dashed lines indicate the start and end times of the second jet. (c) AIA light curves at the jet footpoint (blue box in panel (a)) in the 94, 131, 171, 193, 211, 304, 335, and 1600 $\mathrm{\AA}$ passbands. (d)--(g) four slits (Slice 2,3,4,5) perpendicular to the axis. The two blue dashed lines represent the jet width boundaries, the green stars represent the transverse periodic rotational motion of the jet, and the green dashed lines represent the temporal boundaries of two distinct rotational cycles, whose peak intervals are utilized to compute the rotation period. The two yellow dashed lines indicate the start and end times of the second jet.}
    \label{fig9}%
\end{figure*}

To investigate energy transformation in the jet, we quantify its dynamical evolution through kinematic parameter analysis. By tracking the jet's linear propagation velocity and rotational speed across different altitudes (following the methodology of \citealt{Liu2014}), we derive its mechanical energy. Figure \ref{fig9}(a) illustrates the axial and vertical slits of the jet. Three distinct jet threads are identified in Figure \ref{fig9}(b), which are marked by red stars. Jet thread typically describes fine-scale structures identified within solar jets (\citealt{Schmieder2013}). We acknowledge that all height values used in our energetics calculations are projected heights, as derived from the plane-of-sky observations. This is an inherent limitation when analyzing on-disk events, where the orientation and geometry of the structures intersected by the line of sight are unknown. For the inclined structures investigated in this study (as visible in Figure \ref{fig1}), the projected heights are likely smaller than the actual distances along the inclined magnetic structures. This effect introduces a systematic underestimation in the computed length scales and may lead to an underestimation of the associated energies. While this projection effect does not alter the qualitative trends reported in our analysis, all absolute values of height-dependent parameters should be interpreted as lower limits. We have adopted the term projected height to reflect this uncertainty.

We select three moments 02:49 UT, 02:52 UT, and 02:57 UT to calculate the corresponding velocities of these three jet threads, and the results are shown in Table \ref{table1}. The method for calculating velocity is based on \cite{Liu2009}, \cite{Liu2014}, and \cite{Liu2019}. The axial velocity is derived from the steady propagation phase of the jet thread's motion. For each specified time point, we isolate a segment of the trajectory that exhibits quasi-linear propagation without significant deceleration or acceleration, assuming it represents a period of constant velocity. This segment is fitted with a linear regression in the form of $s(t) = k \cdot t + b$, and the derived slope $k$ denotes the constant velocity for the considered interval. Since the interval is short, this value can be regarded as representative of the speed during that brief period. The error from the fit is taken as the uncertainty in the velocity.

\begin{figure*}
   \centering
   \includegraphics[width=0.5\textwidth]{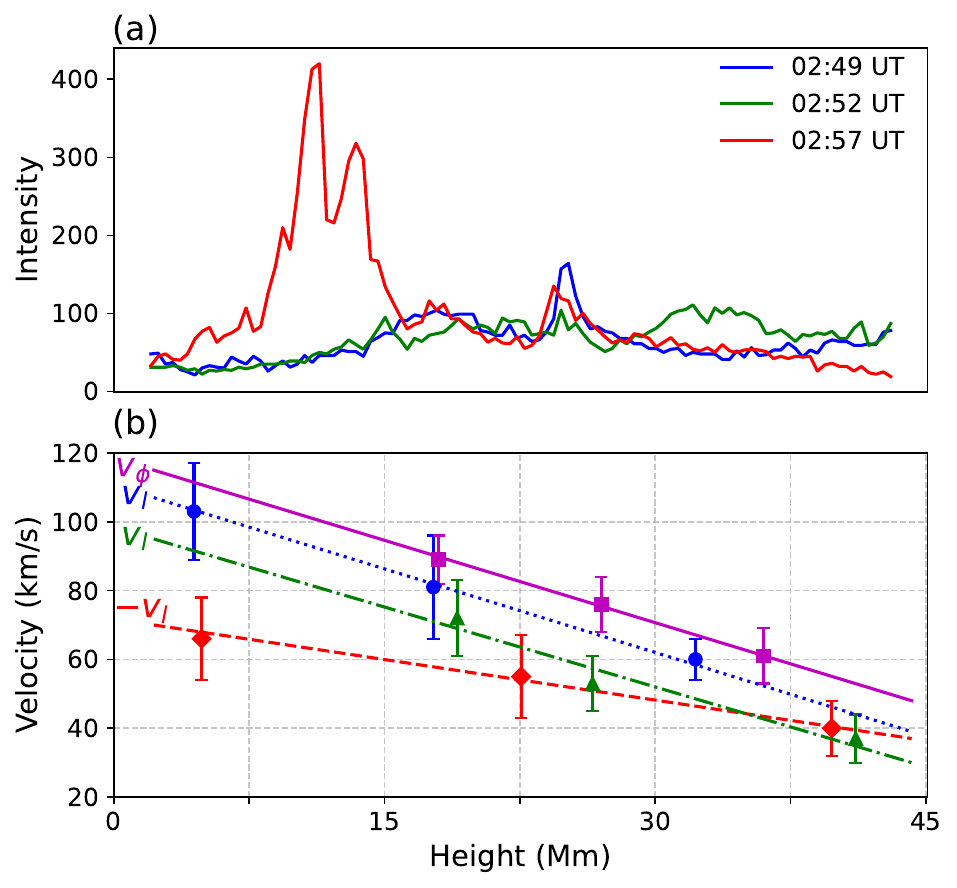}
   \caption{The intensity and velocity of the jet at three moments. (a) 304 $\mathrm{\AA}$ net intensity distribution along projected height at 02:49 UT (blue), 02:52 UT (green), and 02:57 UT (red). (b) Linear velocity $v_l$ and rotation velocity $v_\phi$ distribution along projected height.}
    \label{fig10}%
\end{figure*}

These three moments can represent the initial moment when the jet rises, the moment when the jet reaches its maximum projected height, and the moment when the jet falls back. The transverse slices in Figures \ref{fig9}(d)--(f) reveal periodic rotational patterns across multiple heights. Through tracking the plasma elements (green stars), we observe systematic displacements across the jet's central axis, indicating coherent rotational motion. Our measurements show jet widths (blue dashed lines) and rotation periods (green dashed lines) throughout the structure and the results are shown in Table \ref{table2}. The period is calculated by fitting the times at which the jet material reaches its maximum amplitude (green dashed lines). The time difference between two consecutive instances of reaching the maximum amplitude is defined as the period. The error of the period is derived from the fitting error.
It can be seen that the width of the jet does not change at each projected height. Many previously observed jets produce whiplash motion due to the untwisting of magnetic filed lines, and they become wider with the increase of height (\citealt{Liu2009}; \citealt{Fang2014}; \citealt{Chen2021}). But this jet has no significant broadening, which may indicate that the magnetic structure of the open field lines remains stable for a duration of 5 to 50 minutes. For further testing, we also created multiple slices near the top of the jet, such as slice 5 in Figure \ref{fig9}(g). Similarly, no significant increase in width was observed. However, the slice 5 at the jet's top was not used to derive the rotational speed. This is because the material there is relatively sparse and tenuous, as evidenced by its lower intensity in observations (Figure \ref{fig10}(a)). In cross-sectional slices, sufficiently dense material is required to clearly reveal the rotational signature (as seen in Figures \ref{fig9}(d)--(f)). Near the jet top, the scarcity of material results in a indistinct rotational signature (Figure \ref{fig9}(g)), and any rotational speed measured from this region would be unreliable.

Figure \ref{fig9}(c) displays the AIA light curves in eight passbands at the jet footpoint. The image reveals that all AIA light curves exhibit a primary pulse after 02:40 UT, with its peak indicating the most intense release of magnetic energy through reconnection. This peak coincides precisely with the onset of plasma acceleration at the footpoint (Fig. \ref{fig9}(b)). The subsequent rise of the jet, however, shows a anticipated delay relative to this energy release peak. This delay corresponds to the time required for the accelerated plasma to travel to its maximum height. Shortly after its onset, the jet begins to rotate continuously. In hotter AIA channels such as 94 $\mathrm{\AA}$, 131 $\mathrm{\AA}$, and 335 $\mathrm{\AA}$, the emission peaks after the initial pulse and remains sustained for a period, indicating continuous energy injection at the base of the jet during this interval. After approximately 03:00 UT, a second pulse is observed in the AIA light curves, immediately followed by the occurrence of a second jet. This provides clear evidence for the temporal correspondence between the energy release by magnetic reconnection and the jet events.

Our analysis reveals an intriguing subsequent event following the primary jet eruption, with a recurrent jet emerging at 03:01 UT as shown in Figure \ref{fig9}(b). Although this secondary event's brief duration and incomplete evolutionary cycle (lacking a distinct descending phase) preclude detailed quantitative analysis, its transverse cross-sections shown in Figures \ref{fig9}(d)--(g) demonstrate plasma moves across the central axis. This kinematic pattern suggests that the magnetic structures remain twisted after the first jet. The low plasma density in the second jet likely accounts for the less pronounced rotational signatures compared to the first jet.

Figure \ref{fig10} quantifies the altitude-dependent kinematic characteristics of the jet, with background emission removed by subtracting the average intensity of the pixels around the jet. Both the linear propagation velocity $v_l$ and angular velocity $v_\phi$ exhibit progressive deceleration with increasing projected height—a trend contrasting with the acceleration profiles reported in a rotating jet due to the untwisting of the open magnetic field lines by \cite{Liu2014}. This inverse velocity-altitude relationship suggests fundamentally different driving mechanisms between the two jets. Moreover, the velocities at 02:49 UT and 02:52 UT that are in the ascending phase exhibit a similar distribution pattern with projected height, which may suggest that the magnetic structure remained unchanged throughout the entire jet process.

Following the jet energy calculation method proposed by \cite{Liu2014}, we quantify the jet's mechanical energy components through three fundamental formulae: gravitational potential energy $E_{g}=GM_{\odot}(\int_{0}^{H}I(1/R_{\odot}-1/(h+R_{\odot}))dh/\int_{0}^{H}Idh)$, linear kinetic energy $E_l=\int_0^HIv_{l}^2dh/2\int_0^HIdh$, and rotational kinetic energy $E_{a}=\int_{0}^{H}Iv_{\phi}^{2}dh/2\int_{0}^{H}Idh$. Where $G$ is the gravitational constant, $M_{\odot}$ is the solar mass, $I$ is the intensity at 304 $\mathrm{\AA}$ corresponding to the current height, $H$ is the top height of the jet, $v_l$ is the linear velocity of the jet, and $v_\phi$ is the rotational velocity of the jet. The obtained energy values, with units of J/kg, represent the energy per unit mass. We approximate jet density as proportional to the 304 $\mathrm{\AA}$ intensity after background subtraction. While this is an oversimplification for an optically thick line, which assumes the jet and background are independent layers, it is necessary to avoid severely overestimating the jet's mass and energy by including background plasma emission. Since our analysis focuses on the relative temporal trends of the energies rather than their absolute values, and tests confirm that the trends are unaffected by this subtraction, we consider the method justified.

To address measurement distortions caused by continuous energy injection at the photospheric footpoint, we adopt an innovative fixed-point differential measurement approach.  By analyzing two fixed regions: the base (box1) and apex (box2) of the jet (shown in Figure \ref{fig1}(d)), this method effectively monitors the 'upstream and downstream' mechanical energy of the jet, and the energy changes of the plasma in these two regions can be dynamically displayed. Table \ref{table3} shows the calculated mechanical energy of the two boxes at three moments. Our energy analysis aims to assess the release of free energy from the non-potential magnetic field during the post-reconnection phase. As proposed in previous studies of some jets (e.g., \citealt{Shen2011}; \citealt{Liu2014}), magnetic free energy can be released in two distinct stages: a rapid release via magnetic reconnection (signaled by the peak in the footpoint AIA light curve around 02:40 UT in Figure \ref{fig9}(c)), followed by a more gradual release through the relaxation of twisted magnetic field lines formed during reconnection. The snapshots at 02:49 UT, 02:52 UT, and 02:57 UT were chosen specifically to examine this second stage. It is noteworthy that the initial twist-storing structure differs among these cases: it is the twisted closed field of a small bipolar region in \cite{Shen2011}, a highly twisted magnetic loop in \cite{Liu2014}, and a twisted magnetic flux rope observed as a mini-filament in the present study. However, our shared focus lies in the question of whether the open field subsequently undergoes relaxation and releases energy, after the magnetic reconnection between these initially twisted structures and the background open field transfers the twist to the open field.

It is worth noting that the jet model proposed by \cite{Pariat2009} indicates that magnetic energy injected through slow photospheric rotation can trigger a kink-like instability upon reaching a critical twist, subsequently leading to fast reconnection and nonlinear Alfv\'{e}n waves, which cause a significant increase in kinetic energy during the early stage. The kinetic energy rise described by this model occurs primarily in the initial eruptive phase of jet formation, corresponding to the rapid energy conversion process triggered by reconnection. To characterize the energy evolution during the later development of the jet, our analysis is centered on its motion phase following magnetic reconnection. If significant untwisting were occurring during this phase, it should inject additional mechanical energy, leading to a measurable increase in the total mechanical energy within our observational boxes.

First, comparing observations at 02:49 UT and 02:52 UT reveals three key findings: (1) Box1's gravitational potential energy remains nearly constant, suggesting continuous plasma injection at the jet base during this period. (2) While Box1's linear kinetic energy decreases, its rotational kinetic energy shows minimal change, implying sustained rotational motion at the jet base despite decreasing upward speed during the ascent phase. (3) Box2 exhibits a slight increase in gravitational potential energy with stable linear and rotational kinetic energies, indicating that the jet top maintains rotation while accumulating plasma from continuous bottom supply. Notably, the opposing trends in total mechanical energy (Box1 decreasing vs Box2 increasing) demonstrate energy transfer from base to top during jet ascent. Subsequent analysis of 02:52--02:57 UT shows reversed energy trends: The total mechanical energy of both box1 and box2 keeps decreasing, indicating that the overall mechanical energy of the jet continues to drop. This might be caused by heat energy loss or because some material in the jet spreads out during the downward movement.

Theoretical models and observational evidence indicate that the untwisting of magnetic flux ropes plays a key role in driving energy conversion within jets. Through the release of stored magnetic energy during the untwisting process, this mechanism efficiently converts magnetic energy into both kinetic energy and gravitational potential energy, thereby increasing the system’s total mechanical energy during rotation (\citealt{Shen2011}; \citealt{Liu2014}). In particular, three-dimensional MHD simulations by \cite{Fang2014} demonstrate that jet rotation is dominated by the untwisting motion of post-reconnection magnetic fields, and that this untwisting is central to energy release and transport. This provides a relevant theoretical context for our energy analysis. However, our observations present a contrasting scenario: During the post-reconnection phase covered by our snapshots, both Box1 and Box2 exhibit no significant total mechanical energy increase during jet ascent or descent, which implies the absence of substantial energy injection from magnetic field line relaxation during this stage. This suggests the magnetic structure may persistently maintain its twisted configuration instead of relaxing to a lower-energy state through untwisting. This conclusion aligns with the evolution of the magnetic field structure described in Section \ref{section4.2}. As evident from the comparison between Figures \ref{fig6}(b) and \ref{fig6}(c), the magnetic structure of the external open field maintains its twisted configuration throughout; the field lines do not fully relax into a straightened state. Importantly, this energetics argument should be interpreted with caution. The calculation involves substantial assumptions (e.g., density proxy, background subtraction and the fixed-box method). Therefore, we do not present this as a core or definitive argument but rather as a consistent observational insight that complements the kinematic and magnetic field analyses.

\begin{table}[]
\centering
\caption{The linear velocity $v_l$ and projected height of three jet threads measured at three time points.}
\label{table1}
\begin{tabular}{ccccccc}
\hline
\multirow{3}{*}{Jet threads} & \multicolumn{2}{c}{02:49 UT}    & \multicolumn{2}{c}{02:52 UT}   & \multicolumn{2}{c}{02:57 UT}     \\ \cmidrule(lr){2-3} \cmidrule(lr){4-5} \cmidrule(l){6-7}
                         & $v_l$      & \multicolumn{1}{c}{H} & $v_l$     & \multicolumn{1}{c}{H}  & $v_l$       & \multicolumn{1}{c}{H}  \\
                         & (km/s)   & \multicolumn{1}{c}{(Mm)} & (km/s)  & (Mm) & (km/s)    & \multicolumn{1}{c}{(Mm)} \\ \hline
                         
1                        & 60 $\pm$ 6 & 53.7                    & 37 $\pm$ 7 & 68.5                    & -40 $\pm$ 8  & 66.3                    \\
2                        & 81 $\pm$ 15 & 29.5                    & 53 $\pm$ 8 & 44.2                    & -55 $\pm$ 12 & 37.6                    \\
3                        & 103 $\pm$ 14 & 7.4                     & 72 $\pm$ 11 & 31.7                    & -66 $\pm$ 12 & 8.1                     \\ \hline
\end{tabular}
\end{table}

\begin{table}[]
\centering
\caption{Projected height, width, rotation period, and rotational velocity $v_\phi$ for the jet at three slices.}
\label{table2}
\begin{tabular}{ccccc}
\hline
\multirow{2}{*}{Slice} &Projected height & Width & Period & $v_\phi$     \\
                       & (Mm)     & (Mm)    & (s)      & (km/s)  \\ \hline
2                      & 30     & 7.5  & 264 $\pm$ 20 & 89 $\pm$ 7 \\
3                      & 45     & 7.5  & 310 $\pm$ 32 & 76 $\pm$ 8 \\
4                      & 60     & 7.5  & 384 $\pm$ 49 & 61 $\pm$ 8  \\ \hline
\end{tabular}
\end{table}

\begin{table}[]
\centering
\caption{Gravitational potential energy $E_g$, linear kinetic energy $E_l$, rotational kinetic energy $E_a$, and total mechanical energy $E_t$ of two boxes computed at three time points.}
\label{table3}
\begin{tabular}{ccccccccccccc}
\hline
\multirow{2}{*}{Box} & \multicolumn{4}{c}{02:49 UT} & \multicolumn{4}{c}{02:52 UT} & \multicolumn{4}{c}{02:57 UT} \\

                     & $E_g$    & $E_l$    & $E_a$   & $E_t$    & $E_g$    & $E_l$    & $E_a$    & $E_t$    & $E_g$    & $E_l$    & $E_a$    & $E_t$   \\ \cmidrule(lr){1-1} \cmidrule(lr){2-5} \cmidrule(lr){6-9} \cmidrule(l){10-13}
1                    & 0.37  & 0.39$^{+0.10}_{-0.11}$  & 0.46$^{+0.07}_{-0.05}$  & 1.22$^{+0.17}_{-0.16}$  & 0.36  & 0.30$^{+0.10}_{-0.08}$  & 0.47$^{+0.07}_{-0.06}$  & 1.13$^{+0.17}_{-0.14}$  & 0.32  & 0.20$^{+0.06}_{-0.08}$  & 0.50$^{+0.06}_{-0.06}$  & 1.02$^{+0.12}_{-0.14}$ \\
2                    & 0.90  & 0.14$^{+0.05}_{-0.03}$  & 0.20$^{+0.04}_{-0.04}$  & 1.24$^{+0.09}_{-0.07}$  & 0.94  & 0.10$^{+0.03}_{-0.03}$  & 0.20$^{+0.04}_{-0.04}$  & 1.24$^{+0.07}_{-0.07}$  & 0.91  & 0.10$^{+0.05}_{-0.04}$  & 0.20$^{+0.04}_{-0.04}$  & 1.21$^{+0.09}_{-0.08}$ \\
1+2                  &       &       &       & 2.46$^{+0.26}_{-0.23}$  &       &       &       & 2.37$^{+0.24}_{-0.21}$  &       &       &       & 2.23$^{+0.21}_{-0.22}$ \\ \hline
& & & & & & & & & & & &$\times 10^{10} $ J $\mathrm{kg^{-1}}$
\end{tabular}
\end{table}

\section{Summary and Discussion}\label{section5}
In this study, we analyze a rotating jet and determine the cause of its rotation through combined numerical simulations and observations. The main findings are as follows:

\begin{itemize}

\item[1)]By analyzing CHASE H$\alpha$ and IRIS Si IV spectral lines, we characterize the jet's spectral properties. First, the Doppler velocity reveals predominant redshift on the right and blueshift on the left, consistent with the left-handed spiral rotation of jet material. Second, coexisting redshift and blueshift near the central axis suggest a complex helical magnetic structure within the jet.

\item[2)]Through implementation of  TMF and MHD models, we numerically reproduce the complete process from magnetic flux rope formation to jet eruption, including the rotational dynamics. In the TMF model, a magnetic flux rope consistent with observation forms self-consistently. Subsequently in the MHD model, this structure lifts and reconnects with external open magnetic fields, transferring twist outward. This process enables upward plasma flow along the twisted open field lines, ultimately generating the observed rotational kinematics.

\item[3)]Our study reveals that during the ascending phase, the maximum projected height, and descending phases of the jet, both linear velocity and rotational velocity gradually decrease with increasing height. This contradicts the Sweeping-magnetic-twist acceleration mechanism proposed by \cite{Shibata1986}, indicating that the external open magnetic fields do not untwist to release magnetic energy to accelerate the jet throughout the entire jet process in this case.

\item[4)]Our multi-instrument analysis combining SDO/AIA, CHASE, IRIS observations with MHD simulations reveals the jet's rotation mechanism. Rather than being derived from the external field untwisting, the rotation originates from helical motion of plasma propagating along twisted external magnetic fields. There are two most significant differences between the two mechanisms. One is whether the observed rotation direction of the jet is opposite to the helicity of the magnetic field that threads it. This distinction is critical: rotation opposite to the field's helicity indicates untwisting of the open field as the driver, while rotation in the same direction suggests the plasma is flowing along a helical field. The other is whether the jet material speeds up with altitude or slows down.

\end{itemize}

Based on a synthesis of observational analysis and MHD simulation, we conclude that the rotation of the jet is driven primarily by plasma motion along twisted magnetic field lines, rather than by untwisting of the field lines themselves. This interpretation is robustly supported by multiple lines of evidence: (1) The decrease in both linear and rotational speed with height contradicts the acceleration profile expected from magnetic untwisting. (2) The twisted structure of the open field within the jet remained twisted throughout the jet's lifetime without significant relaxation. (3) The constant jet width across different heights is inconsistent with width expansion with height predicted by the untwisting model. (4) The persistence of transverse plasma motion even during the fallback phase and in a subsequent secondary jet indicates the persistent helical magnetic structure. (5) The observed rotation direction matches the helicity sign of the helical field in the simulation, which is opposite to the direction expected from untwisting. (6) Both the synthetic images and videos clearly reveal plasma rotating along the magnetic field lines. (7) The twist in the open field lines penetrating the jet remained relatively stable throughout the jet process, showing no clear signs of decrease. In summary, these results consistently indicate that the rotation of the jet results from plasma flow along a twisted magnetic structure.

Throughout the process of the rotating jet, the evolution can be divided into three distinct stages from the perspective of magnetic field structure. (1) Helicity transfer: During jet initiation, the small magnetic flux rope at the base rises and undergoes magnetic reconnection with the external open field lines. This transfers its inherent magnetic twist into the external field. In this stage, the observed direction of material rotation should align with the direction of the magnetic field twist. However, this process is extremely rapid, lasting only about 5 minutes in our simulation. (2) Material motion along the twisted structure: During this stage, the external field maintains its twisted configuration. The observed direction of material rotation continues to align with the direction of the magnetic field twist. Compared to the first stage, this phase lasts considerably longer, approximately 45 minutes in our simulation. (3) Untwisting of the magnetic structure: In this final stage, the magnetic field lines of the external field gradually relax and untwist. Theoretically, the observed direction of material rotation should be contrary to the original direction of the magnetic field twist. However, in this specific event, no jet material was produced during the untwisting phase, preventing its direct observation. Previous studies, which interpreted jet rotation as the untwisting motion of the external open field lines and confirmed through observations and simulations that the material motion is directionally contrary to the original twist of the external field, distinctly correspond to this third stage described above. Our study provides a more comprehensive physical explanation for rotating jets, refining their description into these three distinct evolutionary stages.

Many studies have shown that rotating jets typically display spectral features with redshift on one side and blueshift on the other. \cite{Cheung2015} observed spiral motion in a complex jet and used IRIS Si IV spectra to reveal opposite-direction flows at its edges. \cite{Curdt2012} proposed that explosive events observed in transition region spectra are actually narrow rotating jets, with spectral characteristics caused by helical motions in cylindrical structures. \cite{Lee2013} detected cold jets showing spiral motion in He II images, with Doppler redshift and blueshift components supporting rotational mechanisms. However, the one-sided redshift/blueshift pattern only reveals rotation direction and basic spiral structure possibilities. Our findings show isolated redshift/blueshift at jet edges while both shifts coexist along the central axis, suggesting more complex magnetic configurations. These results question whether the simplified rotating cylinder model adequately represents actual jet systems, suggesting we need better models to describe their complex magnetic architectures in future studies.

To investigate whether the observed rotation could be attributed to torsional Alfv\'{e}n waves, we applied cross-correlation to oscillating pairs in adjacent slices from Figures \ref{fig9}(d)--(g) to determine the optimal time difference. By combining these time differences with the corresponding slice heights, we estimated the phase speed of the transverse oscillation to be approximately $v=27$ km $\mathrm{s^{-1}}$. We also estimated the Alfv\'{e}n speed using simulation data. It is noted that the magnetic field strength in the simulation was reduced by a factor of 10 relative to observations to accelerate the calculations (see Section \ref{section3}). For a consistent comparison with the observed phase speed, the field strength used below has been correspondingly scaled back. By averaging the adjusted magnetic field strength $B=9$ G and mass density $\rho=9 \times10^{-14}$ g $\cdot$ $\mathrm{cm^{-3}}$ along magnetic field lines crossing the jet, the resulting Alfv\'{e}n speed is calculated as $v\mathrm{_{A}=\frac{B}{\sqrt{{\mu }_{0}\rho }}}=85$ km $\mathrm{s^{-1}}$. The results show that the phase speed of the transverse oscillation is significantly lower than the Alfv\'{e}n speed. For further validation, we also computed the phase speed of the transverse oscillation in the simulation following the method described in \cite{Lee2015}, by taking transverse slices at different heights of the synthetic 304 $\mathrm{\AA}$ jet images shown in Figure \ref{fig8} and again applying cross-correlation to determine the time difference. This independent approach yielded a consistent result: the phase speed remains well below the local Alfv\'{e}n speed. This indicates that the transverse oscillation of the jet is not driven by Alfv\'{e}n waves, but is more consistent with rotational motion of plasma along twisted magnetic field lines.

The rotational motion in the reported jet likely originates from material flow along twist magnetic field lines. This mechanism of material moving along magnetic structures is common across solar phenomena, such as mass flows in filament eruptions (\citealt{Yan2020}), the Evershed flow in sunspots (\citealt{Evershed1909}; \citealt{Thomas2002}), and plasma flows in spicules (\citealt{Beckers1968}; \citealt{Goodman2012}). This suggests the rotation mechanism observed in jets might extend to the broader solar context, potentially helping us better understand various small-scale motions and other solar phenomena.

\begin{acknowledgements}
We sincerely acknowledge the NASA/SDO science team for providing the AIA and HMI observational data. The numerical computation was conducted in the High Performance Computing Center (HPCC) at Nanjing University. The CHASE mission was successfully launched with support from the China National Space Administration (CNSA). This research was supported by the National Key R\&D Program of China (Grants 2022YFF0503004, 2021YFA1600504, and 2020YFC2201201) and the National Natural Science Foundation of China (Grants 12333009 and 12127901). J.H.G. acknowledges individual support through the China National Postdoctoral Program for Innovative Talents (Grant BX20240159).
\end{acknowledgements}

\bibliography{manu}
\bibliographystyle{aasjournalv7}

\end{document}